\documentclass[preprint]{elsarticle}

\usepackage[version=3]{mhchem} 
\usepackage{amsmath}
\usepackage{amsfonts}
\usepackage{bm}
\usepackage[para]{threeparttable}
\usepackage{graphicx}
\usepackage{caption}
\usepackage{subcaption}
\usepackage{MnSymbol}
\usepackage{color}

\usepackage{lineno,hyperref}
\modulolinenumbers[5]

\journal{Journal of Colloid and Interface Science}

\begin{document}

\begin{frontmatter}

\title{The surface potential explains ion specific bubble coalescence inhibition.}

\author{Timothy T. Duignan\corref{cor1}}
\address{School of Chemical Engineering, The University of Queensland, St Lucia, Brisbane 4072,
Australia}
\ead{t.duignan@uq.edu.au}
\cortext[cor1]{Corresponding author}

\begin{abstract}
\textit{Hypothesis}

\noindent
Some ions can prevent bubbles from coalescing in water. The Gibbs-Marangoni pressure has been proposed as an explanation of this phenomenon. This repulsive pressure occurs during thin film drainage whenever surface enhanced or surface depleted solutes are present. However, bubble coalescence inhibition is known to depend on which particular combination of ions are present in a peculiar and unexplained way. This dependence may be explained by the electrostatic surface potential created by the distribution of ions at the interface, which will alters the natural surface propensity of the ions and hence the Gibbs-Marangoni pressure.

\noindent
\textit{Calculations}

\noindent
A generalised form of the Gibbs-Marangoni pressure is derived for a mixture of solutes and a modified Poisson-Boltzmann equation model is used to calculate this pressure for five different electrolyte solutions made up of four different ions. 

\noindent
\textit{Findings}

\noindent
Combining ions with differing surface propensities, i.e., one enhanced and one depleted, creates a significant electrostatic surface potential which dampens the natural surface propensity of these ions, resulting in a reduced Gibbs-Marangoni pressure, which allows bubble coalescence. This mechanism explains why the ability of electrolytes to inhibit bubble coalescence is correlated with surface tension for pure electrolytes but not for mixed electrolytes.

\end{abstract}


\end{frontmatter}

\section{Introduction}
It has long been known that adding salt to water can inhibit the coalescence of bubbles. This is partly the reason for the white foamy waves that form in seawater but not freshwater. Remarkably, the underlying mechanism of this effect is still not fully understood. This is obviously of fundamental interest but is also crucially important for many practical processes. For example, the mineral extraction technique of flotation relies on this property\cite{Deliyanni2017,Pashley2004}. It can also be used to achieve desalination\cite{Francis2009} and sterilisation.\cite{Sanchis2019}

The standard mechanism proposed to explain the inhibition of bubble coalescence caused by salt was originally described by Marrucci\cite{Marrucci1969}. This is sometimes referred to as the Gibbs-Marangoni elasticity or stress. Here we refer to it as the Gibbs-Marangoni pressure. There has been an ongoing debate for nearly half a century over whether this mechanism can adequately explain the bubble coalescence inhibition effect of salts.\cite{Prince1990,Hofmeier1995,Weissenborn1996,Chan2005,Henry2007,Craig2004,Christenson2008,Craig2011,Horn2011,Firouzi2014,Firouzi2017}

In a recent impressive work, Liu et al. \cite{Liu2020b} give a sophisticated quantitative description of this Gibbs-Marangoni pressure and reproduce the experimentally observed film thickness with no fitted parameters very accurately. This provides strong evidence that this mechanism can adequately explain the ability of salts to inhibit bubble coalescence. 
 
This Gibbs-Marangoni pressure arises due to the fact that as two bubbles come together a thin film drainage process must occur. During this process, there will be a transient change in the concentration of solutes in the thin film. This is because the concentration of solutes in the fluid draining from the film differs from the bulk concentration due to the contribution from the surface region. For surface depleted solutes the concentration in the thin film will increase, whereas for surface enhanced solutes it will decrease. Crucially, according to the Gibbs adsorption isotherm in both these cases this concentration change will act to increase the surface tension in the thin film region. This will temporarily arrest the drainage process via the Marangoni effect thereby preventing coalescence.

This mechanism therefore provides a satisfactory explanation for many of the known properties of bubble coalescence inhibition. Specifically, it occurs for both ions and neutral solutes. It occurs for both surface depleted salts with substantial positive surface tension gradients such as NaCl as well as for surface adsorbed species such as H$\text{ClO}_4$.

However, there remain two critical observations which appear to be entirely qualitatively inconsistent with this mechanism for explaining the bubble coalescence inhibition effect of ions. These observations are:

1) Salts composed of both a surface-enhanced and a surface repelled ion do not appear to inhibit coalescence as  reported by Craig et al.  For example, both Na$\text{ClO}_4$ and HCl do not inhibit coalescence.\cite{Craig1993,Craig1993a} These salts are referred to as $\alpha\beta$  or $\beta\alpha$ where $\alpha$ indicates a surface depleted ion and $\beta$ is a surface-enhanced one.\cite{Marcelja2006,Henry2010} 
When any of these ions are instead paired with ions with matching surface affinity they do inhibit coalescence. For example, NaCl, an $\alpha\alpha$ salt,  and H$\text{ClO}_4$, a $\beta\beta$ salt, both inhibit coalescence. The mechanism of this pairing effect remains unexplained. 

2) Mixtures of $\alpha\beta$ and $\beta\alpha$ salts do inhibit bubble coalescence even though they can have surface tensions almost identical to pure water. The Gibbs-Marangoni pressure relies on a surface tension difference compared to pure water so this observation also appears to be inconsistent with this standard explanation in terms of the Gibbs-Marangoni pressure . \cite{Henry2007}

In this article, I demonstrate that the Gibbs-Marangoni pressure is a consistent explanation of these observations if an additional effect is taken into account. The effect is that ions will create an electrostatic potential at the surface in order to satisfy the electro-neutrality condition. This electrostatic potential will modify the propensity of ions for the interface in a way that alters the Gibbs-Marangoni pressure and explains the observations above.  A modified Poisson-Boltzmann model is used to demonstrate this point quantitatively. 

\section{Theory}

The following expression can be given for the repulsive pressure for one species in solvent due to Gibbs-Marangoni pressure:\cite{Marrucci1969} 

\begin{equation}
P_{\text{GM}}=\frac{4c}{h^2N_\text{A}k_BT}\left(\frac{d\gamma}{dc}\right)^2
\label{eq:PGMpure}
\end{equation}
where $c$ is the concentration of the solute species, $h$ is the thickness of the thin film between the bubbles, $N_\text{A}$ is Avogadro's number,  $k_\text{B}T$ is the Boltzmann constant times temperature and $\gamma$ is the surface tension as a function of concentration of solute. 

This pressure will be counteracted by the Laplace pressure and van der Waals attraction in the bubbles which will drive the bubbles together to rupture. 
\begin{equation}
P_{\text{attr}}=-\frac{2\gamma}{R_\text{bub}}-\frac{B}{h^4}
\label{eq:Pattr}
\end{equation}
where $R_\text{bub}$  is the bubble radius and $B$ is the van der Waals attraction coefficient. 
Potentially other non-DLVO forces may  contribute as well. \cite{Firouzi2017,Ninham2020} 

By equating these pressures, the width of a stable film can be identified and the necessary salt concentration to form a stable film can be identified.\cite{Firouzi2014b} For example, it can be shown that the concentration required to inhibit coalescence should be proportional to $\left( \frac{d\gamma}{dc}\right)^{-2}$.\cite{Prince1990}

In order to understand electrolyte solutions, Eq.~\ref{eq:PGMpure} needs to be generalised for a mixture of solutes. In general, the concentrations of the different solutes can vary independently so a mixture of a surface-active component and a surface depleted component can both give rise independently to a repulsive Gibbs-Marangoni pressure just as they would on their own assuming solute-solute interactions can be neglected, which is reasonable at low concentrations. 

This means that the total surface tension change with total concentration of all solutes is not important. Instead, the total repulsive pressure should just be the sum of the Gibbs-Marangoni pressures caused by each of the solutes in the pure solution. This means the pressure for a mixture can be written as: 
\begin{equation}
P_{\text{GM}}=\frac{4}{h^2N_\text{A}k_\text{B}T}\sum_ic_i\left(\frac{\partial\gamma}{\partial c_i}\right)^2
\label{eq:PGMmix}
\end{equation}
where the sum is over every species ($i$) present. 

It can be unclear what the quantity $\frac{\partial\gamma}{\partial c_i}$ means in the contexts of electrolyte solutions where it is not necessarily possible to experimentally vary the concentration of a single component. It is, therefore, better to express this equation in terms of the surface excess rather than the surface tension. 
These quantities are related through the Gibbs adsorption isotherm 
\begin{equation}
-\frac{1}{N_\text{A}k_\text{B}T}d\gamma=\sum_i\frac{\Gamma_i}{c_i}dc_i
\end{equation}
to give:
\begin{equation}
P_{\text{GM}}=\frac{4N_\text{A}k_\text{B}T}{h^2}\sum_ic_i\left(\frac{\Gamma_i}{c_i}\right)^2
\label{eq:PGMmixSE}
\end{equation}
where $\Gamma_i$ is the surface excess of a given solute:
\begin{equation}
\Gamma_i=\int_{0}^\infty dz \left(c_i(z)-c_i\right)+\int_{-\infty}^0 dz c_i(z)
\label{eq:surfaceEx}
\end{equation}
and where $c_{i}(z)$ is the ionic concentration profile as a function of distance from the surface. A  positive $\Gamma_i$ indicates a strongly surface-enhanced solute which reduces the surface tension and a negative $\Gamma_i $ indicates a surface repelled solute which increases the surface tension. $z=0$ is taken to be the Gibbs dividing surface with $z<0$ giving the gas phase.  $\frac{\Gamma_i}{c_i}$ is typically quite independent of concentration as indicated by the experimentally observed linear surface tension increments.\cite{Henry2007}

These expressions imply that even solutions with a surface tension very close to pure water may still inhibit bubble coalescence if the sum of the squares of the individual surface excesses is sufficiently large. For example, a 50:50  mixture of two neutral species in water. One enhanced and one  depleted may have opposite effects on the surface tension leading to a surface tension that is very similar to bulk water. But bubble coalescence inhibition should still occur for this mixture as both species will independently have a substantial value for the $\left(\frac{\Gamma_i}{c_i}\right)^2$ term. 

These expressions demonstrate that it is not sufficient to know the total surface tension change relative to pure water to predict the Gibbs-Marangoni pressure but instead we must know the surface excesses of the individual  ions. There is no known way to directly measure these quantities for ions. Pegram and Record\cite{Pegram2006} and Marcus\cite{Marcus2010a} have attempted to estimate these quantities by assuming surface tension increments are a linear combination of individual ionic contributions. This assumption is not justifiable as there will be a coupling between the cation and anion surface excesses due to the contribution of the electrostatic potential.  

Therefore, we need to use a theoretical model of this system to determine the surface excesses of the individual ions. The best method for doing this is to use the modified Poisson-Boltzmann  equation given by:
\begin{equation}
-\epsilon_{w} \frac{d^{2} \phi(z)}{dz^{2}}=\sum_{i} q_{i} c_{i}(z)
\end{equation}
where $\epsilon_{w}$ is the dielectric constant of water, $\phi(z)$ is the electrostatic surface potential as a function of distance from the surface, $q_i$ is the ionic charge and the ionic concentration profiles are approximated as: 
\begin{equation}
c_{i}(z)=c_{i}(\infty)\exp{\left[-\beta\left(G_i^{\text{ads}}(z)+q_{i}\phi(z)\right)\right]}
\label{eq:ionDens}
\end{equation}
where $G_i^{\text{ads}}(z)$ is the adsorption free energy of a single solute as a function of distance from the interface. $G_i^{\text{ads}}(z)$ needs to be calculated separately and  input into this equation and can then be numerically solved to determine $\phi(z)$. The inclusion of $G_i^{\text{ads}}(z)$ in the concentration profile makes this a modified Poisson-Boltzmann equation compared to the normal Poisson-Boltzmann equation where this term is ignored. 
 
This is a very widely used and accepted approach to determining ionic distributions and the electrostatic surface potential at interfaces.\cite{Levin2009,Netz2012,Duignan2014b,Uematsu2017,Duignan2018c,Peng2021} Without the inclusion of $G_i^{\text{ads}}(z)$ the effect of ion adsorption on $\gamma$ , $\phi(z)$ and $c_{i}(z)$ cannot be determined. Standard boundary conditions of zero electrostatic field in the vacuum and zero electrostatic potential in the bulk liquid are used to solve the modified Poisson-Boltzmann equation numerically using NDSolve in Mathematica\cite{WolframResearch2012}. Eq.~\ref{eq:ionDens} can be inserted into Eq.~\ref{eq:surfaceEx} which can be inserted into Eq.~\ref{eq:PGMmixSE} to determine the Gibbs-Marangoni pressure. 
 
$G_i^{\text{ads}}(z)$ can be estimated from both molecular simulation methods and continuum solvent models. However, there is a some variation in these predictions depending on which method is chosen. Levin et al.\cite{Levin2009,dosSantos2010} have developed a model for calculating $G_i^{\text{ads}}(z)$ for a number of ions, which we adopt here. This model relies on minimal fitting and satisfactorily reproduces the surface tensions and measured surface potentials of  electrolyte solutions. However, the conclusions of this work are independent of the specific model chosen for  $G_i^{\text{ads}}(z)$ as long as experimental surface tensions are approximately reproduced. To demonstrate this point in the supplementary material we show that the same qualitative conclusions are arrived at if we use a simple box adsorption model for $G_i^{\text{ads}}(z)$ with different parameters that are fitted to reproduce experimental surface tensions. 

\begin{figure}[]
        \centering
	 \includegraphics[width=.6\textwidth]{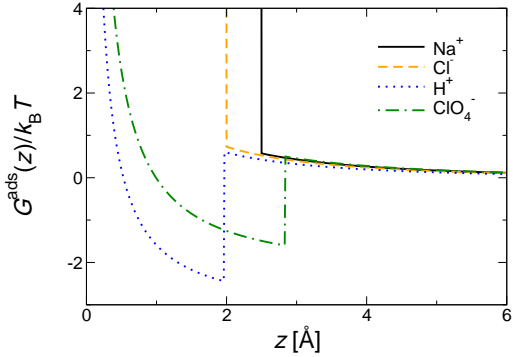}
	     \caption{Adsorption potentials used for the ions.\cite{Levin2009,dosSantos2010}  }
            \label{fig:adsorptionPots}
\end{figure}

In this study we examine solutions composed of combinations of an $\alpha$ cation sodium (Na$^+$), an  $\alpha$ anion chloride (Cl$^-$), a $\beta$ cation  hydrogen (H$^+$) and a $\beta$ anion perchlorate ($\text{ClO}_4$$^-$).  The $G_i^{\text{ads}}(z)$ function for Na$^+$, Cl$^-$, and H$^+$  are shown in Figure~\ref{fig:adsorptionPots} and are exactly the functions developed by Levin et al.\cite{Levin2009,dosSantos2010} The interaction is made up of a long range image charge electrostatic repulsion combined with a hard repulsive wall for Na$^+$ and Cl$^-$  and a square well adsorption for H$^+$. The hard repulsive wall is a standard  approximation for modelling ionic interactions in solution that captures the strong hydration shell of these ions. The adsorption for H$^+$ is qualitatively consistent with more sophisticated analysis.\cite{vacha2008,Duignan2015,Tse2015,Mamatkulov2017}We use these adsorption free energies in this study. However, this method cannot reliably calculate the $G_i^{\text{ads}}(z)$ value for  $\text{ClO}_4^-$ or any other $\beta$ anions. In fact, we are not aware of any reliable calculation of the adsorption free energies of  $\beta$ anions that can reproduce measured surface tensions. For $\text{ClO}_4^-$ we therefore use the same model developed by Levin for the hydrogen ion. For the size parameter we use 2.83 \AA\cite{dosSantos2010} and the depth of the square well adsorption is adjusted to  $-$2.1 $k_\text{B}T$ in order to reproduce experimental surface tensions with reasonable accuracy. A bulk concentration of 0.1 M is used as this is the approximate  concentration range where bubble-bubble coalescence inhibition is observed. 

\section{Results and Discussion}
The resulting surface tension increments, surface tension gradients, surface potentials and Gibbs-Marangoni pressures are given in Table~\ref{table:surfaceprops}. It can be seen that the surface tension gradients are in reasonable agreement with experimental values,\cite{Marcus2010,Henry2007} given the approximate nature of  $G_i^{\text{ads}}(z)$.
\begin{table*}[]
\caption{Surface properties of the electrolytes considered in this study. Experimental surface tension increments are averages\cite{Marcus2010,Henry2007} in units of mN m$^{-1}$ M $^{-1}$. $\frac{\Gamma_-}{c}$ are in units of \AA. The zero of the electrostatic potential is defined to be deep into bulk water in mV. The final column gives the  Gibbs-Marangoni pressure in Pa at 10 nm separation.}
\label{tab:my-table}
\begin{tabular}{c|cccccc}
          & $\frac{\Gamma_+}{c}$  & $\frac{\Gamma_-}{c}$  &$\frac{d\gamma}{dc}$ (Model) & $\frac{d\gamma}{dc}$ (Exp.) & $\phi$ &$P_\text{GM}$ \\ \hline
NaCl     ($\alpha\alpha$)    & $-3.5$                                     & $-3.5$                                     &                1.7                                                                      & 1.8$\pm0.2$                            & -20 &  2400        \\
HCl   ($\beta\alpha$)      & 0.4                                        & 0.4                                        & $-0.2$                                                                                       & $-0.25\pm0.04$                     & 400     &30    \\
$\text{NaClO}_4$  ($\alpha\beta$) & $-1.2$                                     & $-1.2$                                     & 0.6                                                                                         & 0.4  $\pm0.2$                     & -270     &300   \\
$\text{HClO}_4$ ($\beta\beta$)   & 4.2                                        & 4.2                                    & $-2.1$                                                                                       & $-1.7\pm0.6$                    & 150     &3500    \\ \hline
\end{tabular}
\label{table:surfaceprops}
\end{table*}

For NaCl, an $\alpha\alpha$, salt the surface potential is low as the two ions have similar adsorption free energies so there is little charge separation and $\Gamma_+/c=\Gamma_-/c=-3.5 $ \AA. This is a significant surface depletion that gives rise to a substantial Gibbs-Marangoni pressure  as shown in the final column explaining why these salts inhibit coalescence. In contrast, for the $\beta\alpha$ salt HCl, a substantial positive surface potential develops.  This positive surface potential is caused by the adsorption of hydrogen ions to the interface\cite{Duignan2015} combined with the chloride ion repulsion, which creates a charge separation. This positive surface potential attracts chloride ions to the surface mostly cancelling out their natural propensity to be repelled from the interface. Similarly, it repels hydrogen ions reducing their natural enhancement at the interface. This results in a minimal Gibbs-Marangoni pressure that is more than an order of magnitude smaller than for NaCl explaining why HCl does not inhibit coalescence.  A similar but sign reversed mechanism occurs for Na$\text{ClO}_4$ again resulting in minimal Gibbs-Marangoni pressure. Notably, these values are very different from what would be expected if the surface potential were neglected as seen by the value of  $\frac{\Gamma_-}{c}$ for Cl$^{-}$ in  NaCl solution where there is little surface potential.  For the $\beta\beta$ salt, H$\text{ClO}_4$, a relatively small surface potential and large surface excess occurs for both ions and hence a large Gibbs-Marangoni pressure results. This model therefore provides an explanation of the first observation associated with the ion-specificity of bubble coalescence inhibition. Interestingly, the Gibbs-Marangoni pressure term, at the low concentrations studied here, for the  $\alpha\beta$  and $\beta\alpha$  electrolytes is not exactly zero as the surface excesses are not precisely zero. This theory therefore predicts that at sufficiently high concentrations even these ions should eventually be able to inhibit coalescence. This has in fact been observed experimentally.\cite{Christenson2008} However, at these  higher concentrations various assumptions of the model can begin to break down such as linear surface tensions and activity coefficients of 1. These effects would need to be accounted for in order to make quantitative conclusions. 

\begin{figure*}[]
        \centering
        \begin{subfigure}{.49\textwidth}
	 \includegraphics[width=1\textwidth]{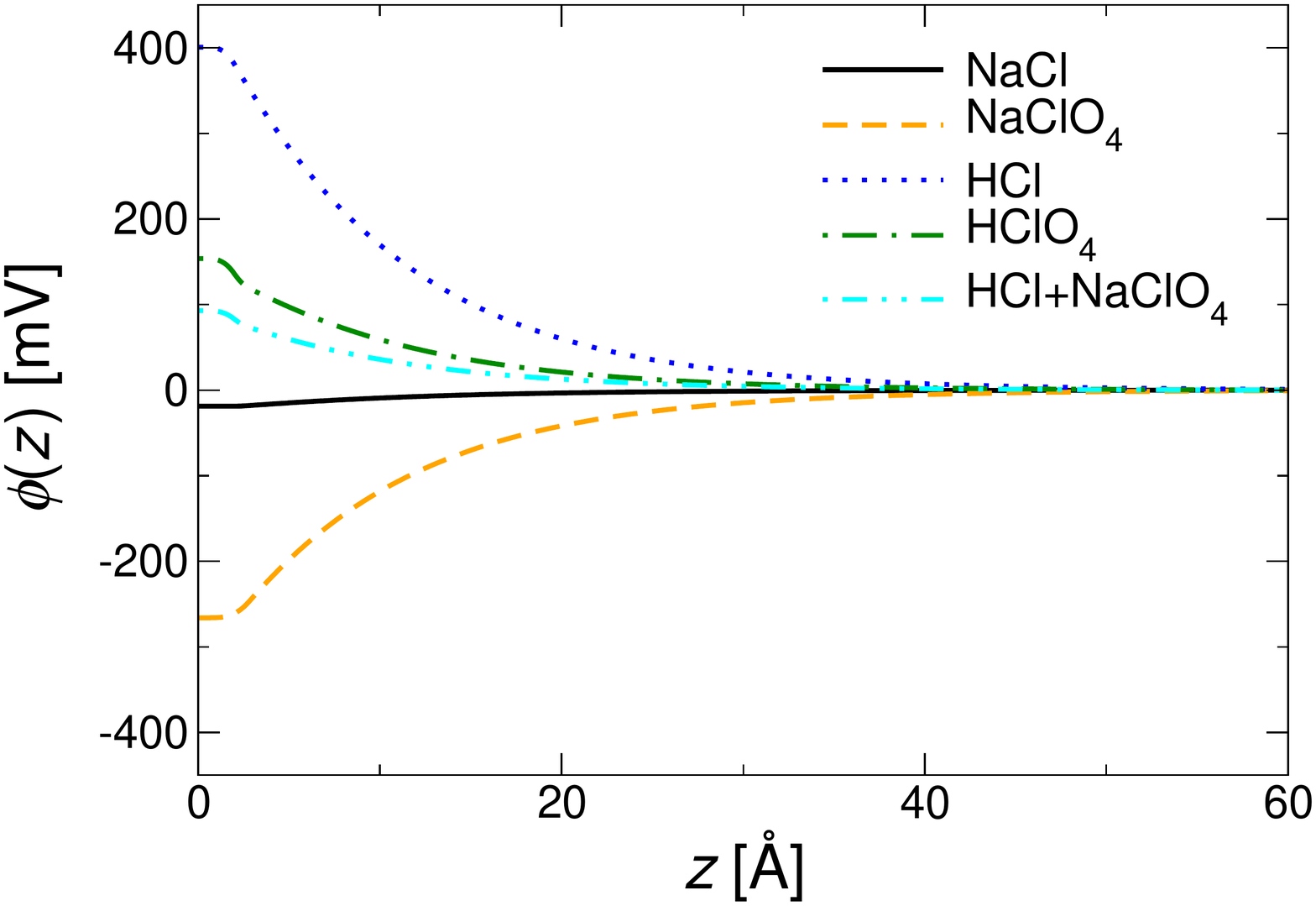}
	     \caption{}
	       \label{fig:surfacepota}
       \end{subfigure}
       \begin{subfigure}{.49\textwidth}
	 \includegraphics[width=1\textwidth]{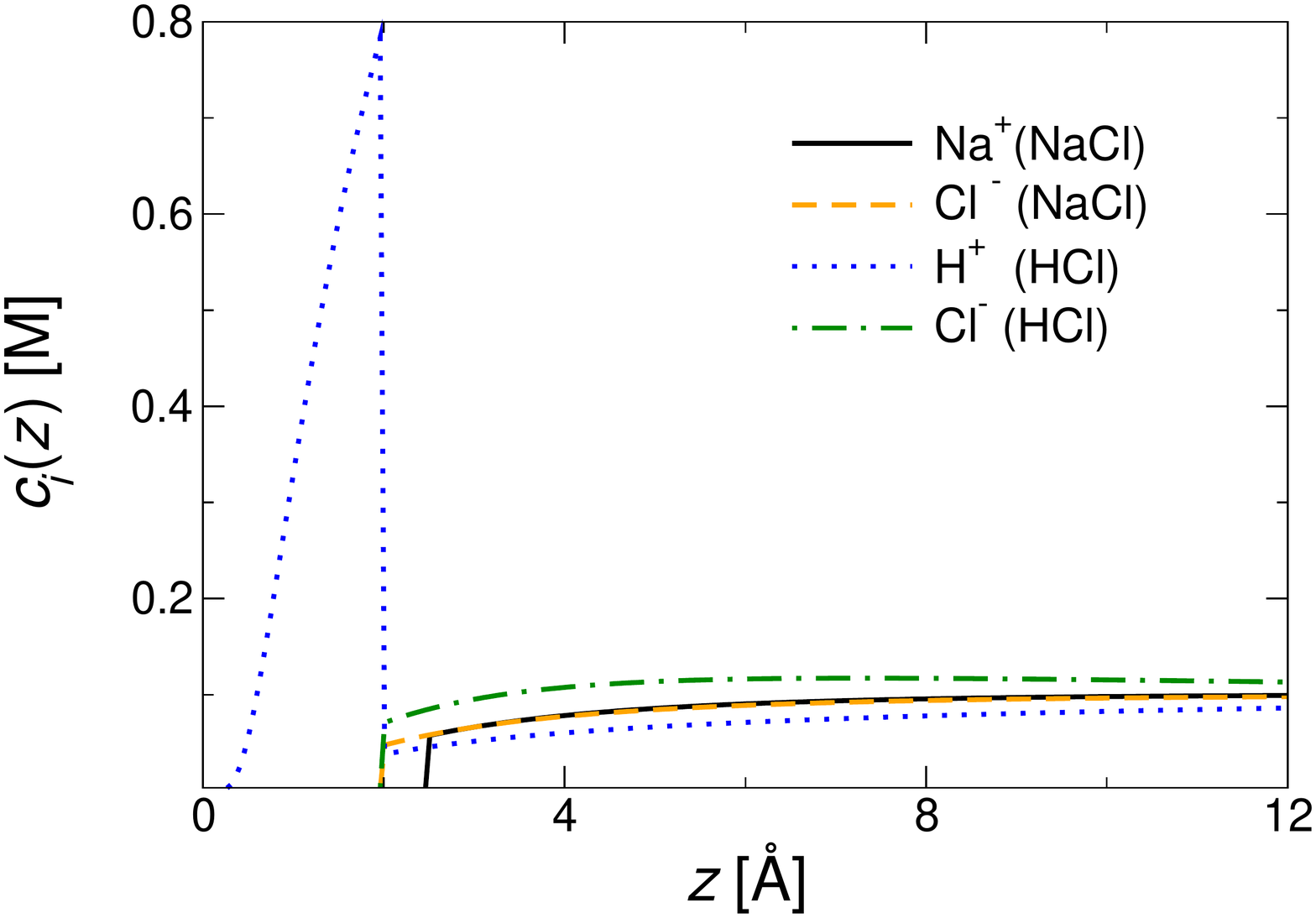}
	 \caption{}
	  \label{fig:surfacepotb}
          \end{subfigure}
	               \caption[]{(a) Surface potentials for the 4 electrolyte solutions studied here as well as one electrolyte mixture. (b) Ionic density profiles for two electrolyte solutions.  }
            \label{fig:surfacepot}
\end{figure*}

Figure~\ref{fig:surfacepota} shows the surface potential for the five cases examined here showing the large positive and negative potential for the $\beta\alpha$ and $\alpha\beta$ salts. Figure~\ref{fig:surfacepotb} shows the impact of this surface potential on the density profiles of the ions. For HCl the surface potential creates a long-range depletion of H$^+$ and enhancement of Cl$^-$ which cancels their short-range surface excesses. 

All of the pure electrolytes satisfy the electro-neutrality condition as they must, i.e.,  $\Gamma_+=\Gamma_-$. This is a result of the boundary conditions used in solving the modified Poisson-Boltzmann equation. In fact, the surface potential is the mechanism by which electro-neutrality of the interface is enforced. From the Gibbs adsoprtion isotherm, the  electro-neutrality condition also gives rise to the following useful expression for the contribution of individual ions to the total surface tension for pure monovalent salt mixtures. 
\begin{equation} \frac{\partial\gamma}{\partial c_+}=-N_\text{A}k_\text{B}T\frac{\Gamma_+}{c_+}=-N_\text{A}k_\text{B}T\frac{\Gamma_-}{c_-}=\frac{\partial\gamma}{\partial c_-}=\frac{1}{2}\left(\frac{d\gamma}{dc}\right)
\label{eq:ENcond}
\end{equation}

This expression means that for pure electrolytes, made up of only one cation-anion pair, the surface excess of each individual ion can be straightforwardly determined from the total surface tension gradients. This explains why the transition concentration where bubble coalescence inhibition occurs can be related to the measured total surface tension gradient.\cite{Prince1990,Yaminsky1995,Firouzi2017} Similar expressions can be derived for pure electrolytes with multivalent ions.

However, this simple expression for the surface excess is not applicable to mixtures of electrolyte solutions, as the surface charge associated with the surface excess of one ion can be cancelled by a combination of the three other ion's surface excesses.  This explains why the simple correlation of transition concentration and surface tension gradient breaks down for electrolyte mixtures.\cite{Henry2007} 

For electrolyte mixtures the surface potential will depend on specifically which electrolytes are mixed.  When $\alpha\beta$ electrolytes are mixed with other $\alpha\beta$ electrolytes the surface potential will still be substantial as the surface potentials of these electrolytes will approximately add together. This explains why these mixtures also do not inhibit coalescence experimentally.\cite{Henry2007} Similarly, for $\beta\alpha$ electrolytes mixed with other $\beta\alpha$ electrolytes. However, when $\alpha\beta$ and $\beta\alpha$ salts are mixed the surface potential of these two electrolytes will cancel as they have opposite signs allowing the natural propensity of the ions for the interface to emerge again leading to bubble coalescence inhibition. In other words, electro-neutrality can be satisfied because the excess anions at the surface of the $\alpha\beta$ will be cancelled by the excess of cations of the $\beta\alpha$ salt. 

\begin{table*}[]
\caption{Surface properties of the mixed electrolyte. The surface tension increments are in units of mN m$^{-1}$ M $^{-1}$. Experimental error is an average for other salts. $\frac{\Gamma_-}{c}$ are in units of \AA.  The final column gives the Gibbs-Marangoni pressure.}
\label{tab:my-table}
\begin{tabular}{c|cccccccc}
          & $\frac{\Gamma_{\text{Na}^+}}{c}$  & $\frac{\Gamma_\text{Cl$^-$}}{c}$& $\frac{\Gamma_\text{H$^+$}}{c}$& $\frac{\Gamma_{\text{ClO}_4^-}}{c}$  &$\frac{d\gamma}{dc}$ (Model) & $\frac{d\gamma}{dc}$ (Exp.)\cite{Henry2007} & $\phi$ (mV)&$P_\text{GM}$ \\ \hline
 HCl  +Na$\text{ClO}_4$   & $-4.3$                                     & $-2.6$                                  &                 5.1      &        3.4                     &              $-0.2$                           & -0.2$\pm0.2$                      & 90 &  3100       \\
\hline
\end{tabular}
\label{table:surfacepropsmix}
\end{table*}

This case can also be modelled with the modified Poisson-Boltzmann equation. Table~\ref{table:surfacepropsmix} gives the relevant values. From comparison with Table~\ref{table:surfaceprops} it is clear that the mixture behaves much more similarly to an $\alpha\alpha$ or $\beta\beta$ salt than to an $\alpha\beta$ salt. As it has a smaller surface potential leading to large values for the individual ion excesses and a large contribution to the Gibbs-Marangoni pressure. (Note the concentrations of each ion are halved to determine the Gibbs-Marangoni pressure comparable with pure electrolyte.) The individual surface excesses are not constrained by the electro-neutrality condition to be equal so there is a large cancellation resulting in minimal surface tension gradient consistent with experiment. It should be noted that the sum of cation excesses is still equal to the sum of anion excesses as must be the case to satisfy electroneutrality overall. This model therefore also provides an explanation of the second observation associated with the ion-specificity of bubble coalescence inhibition, which is that certain salt mixtures such as Na$\text{ClO}_4$ and HCl have surface tensions very similar to water but bubble coalescence inhibition does occur. 

Using this model it should be possible to provide quantitative predictions of the transition concentration of all electrolytes. However, to achieve this the model will need to be coupled with a full understanding and more sophisticated modelling of several additional aspects that may play a role in determining the transition concentration. Specifically, 1) the effect of diffusion of ions, which will cause the film to narrow at the edges 2) the effect of fluctuations such as capillary waves on the surface of the bubbles, which may cause narrowing of the thin film at points\cite{Ruckenstein1974} 3) the effect of other neglected surface forces potentially associated with dissolved gas or the distortion of the bubble surface\cite{Firouzi2017,Ninham2020}  and 4) The effect of the streaming potential and mobility of the film surfaces \cite{Tsekov2010,Karakashev2011}This will need to be combined with a precise determination of the various parameters required in the equations such as surface tensions and bubble radius, which can be very difficult to obtain precisely and is beyond the scope of this work.  

\section{Conclusion}
In summary, this paper provides a new expression for the Gibbs-Marangoni pressure for solute mixtures. It combines this equation with the modified Poisson-Boltzmann equation to calculate the Gibbs-Marangoni pressure for five different electrolyte solutions. The resulting key insight is that when ions with opposite surface propensity are combined a significant surface potential is created which acts to negate the natural surface propensity of the ions minimising their Gibbs-Marangoni pressure and explaining why these electrolytes do not inhibit bubble coalescence. This verifies the hypothesis that the surface potential is crucial to explaining the ion specific behaviour of bubble coalescence. 

Crucially, this explanation accounts for why mixed electrolytes with minimal surface tension gradient can still inhibit bubble coalescence. No satisfactory explanation of this observation has previously been proposed and was until now believed  to rule out the Gibbs-Marangoni pressure as the possible mechanism of bubble coalescence inhibition. \cite{Henry2007}

Only one explanation of the ion specificity of bubble coalescence inhibition has previously been proposed.\cite{Katsir2014a} This explanation assumes that a charge separation between ions forms at the interface leading to an electrostatic electric double layer repulsion which prevents coalescence. While it is correct that the charge separation of ions at the interface is the key to understanding this phenomenon, this explanation actually implies precisely the opposite ion specific behaviour to what is actually observed. Specifically, $\alpha\beta$ and $\beta\alpha$ salts have the largest charge separation and should therefore have the largest electrostatic repulsion and inhibit coalescence most strongly,  but these are exactly the salts that do not inhibit coalescence.\cite{Henry2010} Additionally, the concentrations at which this phenomenon occur mean that the electrostatic repulsion is screened too strongly for this to be the correct explanation. The origin of this error arises from the view that hydroxide strongly adsorbs to the air water interface. A significant body of evidence has now accumulated indicating that this is not the case.\cite{vacha2008,Mundy2009,Baer2014a,Duignan2015,Tse2015,Uematsu2018b}

This means that the traditional explanation of bubble coalescence inhibition in terms of the Gibbs-Marangoni pressure originally proposed by Marrucci\cite{Marrucci1969} can be regarded as correct, resolving a 50 year old debate on this topic. Determining the fundamental forces that drive interactions of particles and bubbles in solution is obviously of central importance to colloid and interface science and the Gibbs-Marangoni pressure is likely to arise in any situation involving a thin-film drainage process in the presence of surface enhanced or depleted solutes. In fact it is one of the fundamental mechanisms that explains the stability of colloidal systems\cite{SweetaAkbari2018} so its accurate description is of broad significance. Now a  quantitive description of this pressure for mixed electrolytes is possible. But even more generally the distribution of ions at surfaces play an important role in determining many important interfacial and colloidal properties such as zeta potential, surface tension and reaction rates. This work provides a demonstration of how to accurately account for the role of the surface potential created by the ions on these properties. 

In future this work should be combined with an accurate description of the attractive forces between bubbles to provide a precise quantitative predictions of the critical salt concentrations required to inhibit bubble coalescence in mixed electrolytes.\cite{Firouzi2017}  The theory should also be applied to the many practical cases where these bubble solutions are practically important. For instance, predicting the conditions of bubble adhesion to mineral particles is a problem of immense practical importance in industry where it is at the heart of  flotation process for purifying minerals.\cite{Deliyanni2017,Pashley2004} 

\section{Acknowledgements}
TTD acknowledge the Australian Research Council (ARC) funding via project number DE200100794 and DP200102573. Thank you to Vince Craig, Barry Ninham and Drew Parsons for helpful discussion.

\section*{References}
\bibliography{library}

\begin{thebibliography}{10}
\expandafter\ifx\csname url\endcsname\relax
  \def\url#1{\texttt{#1}}\fi
\expandafter\ifx\csname urlprefix\endcsname\relax\def\urlprefix{URL }\fi
\expandafter\ifx\csname href\endcsname\relax
  \def\href#1#2{#2} \def\path#1{#1}\fi

\bibitem{Deliyanni2017}
E.~A. Deliyanni, G.~Z. Kyzas, K.~A. Matis,
  \href{http://linkinghub.elsevier.com/retrieve/pii/S0167732216332482}{{Various
  flotation techniques for metal ions removal}}, J. Mol. Liq. 225 (2017)
  260--264.
\newblock \href {http://dx.doi.org/10.1016/j.molliq.2016.11.069}
  {\path{doi:10.1016/j.molliq.2016.11.069}}.
\newline\urlprefix\url{http://linkinghub.elsevier.com/retrieve/pii/S0167732216332482}

\bibitem{Pashley2004}
R.~M. Pashley, M.~E. Karaman, {Applied Colloid and Surface Chemistry}, John
  Wiley \& Sons, Ltd, 2004.

\bibitem{Francis2009}
M.~J. Francis, R.~M. Pashley,
  \href{http://www.tandfonline.com/doi/abs/10.5004/dwt.2009.917}{{Thermal
  desalination using a non-boiling bubble column}}, Desalin. Water Treat.
  12~(1-3) (2009) 155--161.
\newblock \href {http://dx.doi.org/10.5004/dwt.2009.917}
  {\path{doi:10.5004/dwt.2009.917}}.
\newline\urlprefix\url{http://www.tandfonline.com/doi/abs/10.5004/dwt.2009.917}

\bibitem{Sanchis2019}
A.~G. Sanchis, R.~Pashley, B.~Ninham,
  \href{http://dx.doi.org/10.1038/s41545-018-0027-5}{{Virus and bacteria
  inactivation by CO2 bubbles in solution}}, npj Clean Water 2 (2019) 5.
\newblock \href {http://dx.doi.org/10.1038/s41545-018-0027-5}
  {\path{doi:10.1038/s41545-018-0027-5}}.
\newline\urlprefix\url{http://dx.doi.org/10.1038/s41545-018-0027-5}

\bibitem{Marrucci1969}
G.~Marrucci, {A theory of coalescence}, Chem. Eng. Sci. 24~(6) (1969) 975--985.
\newblock \href {http://dx.doi.org/10.1016/0009-2509(69)87006-5}
  {\path{doi:10.1016/0009-2509(69)87006-5}}.

\bibitem{Prince1990}
M.~J. Prince, H.~W. Blanch, {Transition electrolyte concentrations for bubble
  coalescence}, AlChe 36~(9) (1990) 1425--1429.
\newblock \href {http://dx.doi.org/10.1002/aic.690360915Citat}
  {\path{doi:10.1002/aic.690360915Citat}}.

\bibitem{Hofmeier1995}
U.~Hofmeier, V.~V. Yaminsky, H.~K. Christenson, {Observations of solute effects
  on bubble formation}, J. Colloid Interface Sci. 174~(1) (1995) 199--210.
\newblock \href {http://dx.doi.org/10.1006/jcis.1995.1383}
  {\path{doi:10.1006/jcis.1995.1383}}.

\bibitem{Weissenborn1996}
P.~K. Weissenborn, R.~J. Pugh, {Surface Tension of Aqueous Solutions of
  Electrolytes: Relationship with Ion Hydration, Oxygen Solubility, and Bubble
  Coalescence}, J. Colloid Interface Sci. 184 (1996) 550--563.
\newblock \href {http://dx.doi.org/10.1006/jcis.1996.0651}
  {\path{doi:10.1006/jcis.1996.0651}}.

\bibitem{Chan2005}
B.~S. Chan, Y.~H. Tsang, {A theory on bubble-size dependence of the critical
  electrolyte concentration for inhibition of coalescence}, J. Colloid
  Interface Sci. 286 (2005) 410--413.
\newblock \href {http://dx.doi.org/10.1016/j.jcis.2004.01.026}
  {\path{doi:10.1016/j.jcis.2004.01.026}}.

\bibitem{Henry2007}
C.~L. Henry, C.~N. Dalton, L.~Scruton, V.~S.~J. Craig,
  \href{http://pubs.acs.org/doi/abs/10.1021/jp066400b}{{Ion-Specific
  Coalescence of Bubbles in Mixed Electrolyte Solutions}}, J. Phys. Chem. C
  111~(2) (2007) 1015--1023.
\newblock \href {http://dx.doi.org/10.1021/jp066400b}
  {\path{doi:10.1021/jp066400b}}.
\newline\urlprefix\url{http://pubs.acs.org/doi/abs/10.1021/jp066400b}

\bibitem{Craig2004}
V.~Craig,
  \href{http://linkinghub.elsevier.com/retrieve/pii/S1359029404000627}{{Bubble
  coalescence and specific-ion effects}}, Curr. Opin. Colloid Interface Sci.
  9~(1-2) (2004) 178--184.
\newblock \href {http://dx.doi.org/10.1016/j.cocis.2004.06.002}
  {\path{doi:10.1016/j.cocis.2004.06.002}}.
\newline\urlprefix\url{http://linkinghub.elsevier.com/retrieve/pii/S1359029404000627}

\bibitem{Christenson2008}
H.~K. Christenson, R.~E. Bowen, J.~A. Carlton, J.~R. Denne, Y.~Lu,
  {Electrolytes that show a transition to bubble coalescence inhibition at high
  concentrations}, J. Phys. Chem. C 112~(3) (2008) 794--796.
\newblock \href {http://dx.doi.org/10.1021/jp075440s}
  {\path{doi:10.1021/jp075440s}}.

\bibitem{Craig2011}
V.~S.~J. Craig, \href{http://dx.doi.org/10.1016/j.cocis.2011.04.003
  http://linkinghub.elsevier.com/retrieve/pii/S135902941100046X}{{Do hydration
  forces play a role in thin film drainage and rupture observed in electrolyte
  solutions?}}, Curr. Opin. Colloid Interface Sci. 16~(6) (2011) 597--600.
\newblock \href {http://dx.doi.org/10.1016/j.cocis.2011.04.003}
  {\path{doi:10.1016/j.cocis.2011.04.003}}.
\newline\urlprefix\url{http://dx.doi.org/10.1016/j.cocis.2011.04.003
  http://linkinghub.elsevier.com/retrieve/pii/S135902941100046X}

\bibitem{Horn2011}
R.~G. Horn, L.~A. {Del Castillo}, S.~Ohnishi,
  \href{http://dx.doi.org/10.1016/j.cis.2011.05.006}{{Coalescence map for
  bubbles in surfactant-free aqueous electrolyte solutions}}, Adv. Colloid
  Interface Sci. 168~(1-2) (2011) 85--92.
\newblock \href {http://dx.doi.org/10.1016/j.cis.2011.05.006}
  {\path{doi:10.1016/j.cis.2011.05.006}}.
\newline\urlprefix\url{http://dx.doi.org/10.1016/j.cis.2011.05.006}

\bibitem{Firouzi2014}
M.~Firouzi, T.~Howes, A.~V. Nguyen,
  \href{http://linkinghub.elsevier.com/retrieve/pii/S0001868614002292}{{A
  quantitative review of the transition salt concentration for inhibiting
  bubble coalescence}}, Adv. Colloid Interface Sci. 222 (2015) 305--318.
\newblock \href {http://dx.doi.org/10.1016/j.cis.2014.07.005}
  {\path{doi:10.1016/j.cis.2014.07.005}}.
\newline\urlprefix\url{http://linkinghub.elsevier.com/retrieve/pii/S0001868614002292}

\bibitem{Firouzi2017}
M.~Firouzi, A.~V. Nguyen,
  \href{http://dx.doi.org/10.1016/j.colsurfa.2016.12.004}{{The Gibbs-Marangoni
  stress and nonDLVO forces are equally important for modeling bubble
  coalescence in salt solutions}}, Colloids Surfaces A Physicochem. Eng. Asp.
  515 (2017) 62--68.
\newblock \href {http://dx.doi.org/10.1016/j.colsurfa.2016.12.004}
  {\path{doi:10.1016/j.colsurfa.2016.12.004}}.
\newline\urlprefix\url{http://dx.doi.org/10.1016/j.colsurfa.2016.12.004}

\bibitem{Liu2020b}
B.~Liu, R.~Manica, Z.~Xu, Q.~Liu, \href{http://arxiv.org/abs/2007.10972}{{Ion
  specificity modulated inhomogeneous interfacial flow inhibits bubble
  coalescence in electrolyte solutions}}, arXiv\href
  {http://arxiv.org/abs/2007.10972} {\path{arXiv:2007.10972}}.
\newline\urlprefix\url{http://arxiv.org/abs/2007.10972}

\bibitem{Craig1993}
V.~S.~J. Craig, B.~W. Ninham, R.~M. Pashley, {Effect of Electrolytes on Bubble
  Coalescence}, Nature 364~(6435) (1993) 317--319.
\newblock \href {http://dx.doi.org/10.1038/364317a0}
  {\path{doi:10.1038/364317a0}}.

\bibitem{Craig1993a}
V.~S.~J. Craig, B.~W. Ninham, R.~M. Pashley, {The Effect of Electrolytes on
  Bubble Coalescence in Water}, J. Phys. Chem. 97~(type 2) (1993) 10192--10197.
\newblock \href {http://dx.doi.org/10.1021/j100141a047}
  {\path{doi:10.1021/j100141a047}}.

\bibitem{Marcelja2006}
S.~Marcelja, {Selective Coalescence of Bubbles in Simple Electrolytes}, J.
  Phys. Chem. B 110 (2006) 13062--13067.
\newblock \href {http://dx.doi.org/10.1021/jp0610158}
  {\path{doi:10.1021/jp0610158}}.

\bibitem{Henry2010}
C.~L. Henry, V.~S.~J. Craig,
  \href{http://www.ncbi.nlm.nih.gov/pubmed/20092343}{{The Link between Ion
  Specific Bubble Coalescence and Hofmeister Effects is the Partitioning of
  Ions within the Interface.}}, Langmuir 26~(9) (2010) 6478--6483.
\newblock \href {http://dx.doi.org/10.1021/la9039495}
  {\path{doi:10.1021/la9039495}}.
\newline\urlprefix\url{http://www.ncbi.nlm.nih.gov/pubmed/20092343}

\bibitem{Ninham2020}
B.~W. Ninham, P.~L. Nostro, {Unexpected Properties of Degassed Solutions
  Unexpected Properties of Degassed Solutions}, J. Phys. Chem. B 124 (2020)
  7872--7878.
\newblock \href {http://dx.doi.org/10.1021/acs.jpcb.0c05001}
  {\path{doi:10.1021/acs.jpcb.0c05001}}.

\bibitem{Firouzi2014b}
M.~Firouzi, A.~V. Nguyen, {Novel methodology for predicting the critical salt
  concentration of bubble coalescence inhibition}, J. Phys. Chem. C 118~(2)
  (2014) 1021--1026.
\newblock \href {http://dx.doi.org/10.1021/jp409473g}
  {\path{doi:10.1021/jp409473g}}.

\bibitem{Pegram2006}
L.~M. Pegram, M.~T. Record,
  \href{http://www.pubmedcentral.nih.gov/articlerender.fcgi?artid=1599954&tool=pmcentrez&rendertype=abstract}{{Partitioning
  of atmospherically relevant ions between bulk water and the water/vapor
  interface.}}, Proc. Natl. Acad. Sci. U. S. A. 103~(39) (2006) 14278--81.
\newblock \href {http://dx.doi.org/10.1073/pnas.0606256103}
  {\path{doi:10.1073/pnas.0606256103}}.
\newline\urlprefix\url{http://www.pubmedcentral.nih.gov/articlerender.fcgi?artid=1599954&tool=pmcentrez&rendertype=abstract}

\bibitem{Marcus2010a}
Y.~Marcus, \href{http://pubs.acs.org/doi/abs/10.1021/je1002175}{{Surface
  Tension of Aqueous Electrolytes and Ions}}, J. Chem. Eng. Data 55~(9) (2010)
  3641--3644.
\newblock \href {http://dx.doi.org/10.1021/je1002175}
  {\path{doi:10.1021/je1002175}}.
\newline\urlprefix\url{http://pubs.acs.org/doi/abs/10.1021/je1002175}

\bibitem{Levin2009}
Y.~Levin, A.~P. dos Santos, A.~Diehl,
  \href{http://link.aps.org/doi/10.1103/PhysRevLett.103.257802}{{Ions at the
  Air-Water Interface: An End to a Hundred-Year-Old Mystery?}}, Phys. Rev.
  Lett. 103~(25) (2009) 257802.
\newblock \href {http://dx.doi.org/10.1103/PhysRevLett.103.257802}
  {\path{doi:10.1103/PhysRevLett.103.257802}}.
\newline\urlprefix\url{http://link.aps.org/doi/10.1103/PhysRevLett.103.257802}

\bibitem{Netz2012}
R.~R. Netz, D.~Horinek,
  \href{http://www.ncbi.nlm.nih.gov/pubmed/22404593}{{Progress in Modeling of
  Ion Effects at the Vapor/Water Interface.}}, Annu. Rev. Phys. Chem. 63 (2012)
  401--418.
\newblock \href {http://dx.doi.org/10.1146/annurev-physchem-032511-143813}
  {\path{doi:10.1146/annurev-physchem-032511-143813}}.
\newline\urlprefix\url{http://www.ncbi.nlm.nih.gov/pubmed/22404593}

\bibitem{Duignan2014b}
T.~T. Duignan, D.~F. Parsons, B.~W. Ninham, {Ion interactions with the
  air-water interface using a continuum solvent model}, J. Phys. Chem. B
  118~(29) (2014) 8700--8710.
\newblock \href {http://dx.doi.org/dx.doi.org/10.1021/jp502887e}
  {\path{doi:dx.doi.org/10.1021/jp502887e}}.

\bibitem{Uematsu2017}
Y.~Uematsu, D.~J. Bonthuis, R.~R. Netz,
  \href{http://pubs.acs.org/doi/10.1021/acs.jpclett.7b02960}{{Charged
  surface-active impurities at nanomolar concentration induce Jones-Ray
  effect}}, J. Phys. Chem. Lett. 9~(1) (2018) 189--193.
\newblock \href {http://dx.doi.org/10.1021/acs.jpclett.7b02960}
  {\path{doi:10.1021/acs.jpclett.7b02960}}.
\newline\urlprefix\url{http://pubs.acs.org/doi/10.1021/acs.jpclett.7b02960}

\bibitem{Duignan2018c}
T.~T. Duignan, M.~Peng, A.~V. Nguyen, X.~S. Zhao, M.~D. Baer, C.~J. Mundy,
  \href{http://aip.scitation.org/doi/10.1063/1.5050421}{{Detecting the
  undetectable: The role of trace surfactant in the Jones-Ray effect}}, J.
  Chem. Phys. 149~(19) (2018) 194702.
\newblock \href {http://dx.doi.org/10.1063/1.5050421}
  {\path{doi:10.1063/1.5050421}}.
\newline\urlprefix\url{http://aip.scitation.org/doi/10.1063/1.5050421}

\bibitem{Peng2021}
M.~Peng, T.~T. Duignan, C.~V. Nguyen, A.~V. Nguyen, {From Surface Tension to
  Molecular Distribution: Modeling Surfactant Adsorption at the Air - Water
  Interface}, Langmuir 37~(7) (2021) 2237--2255.
\newblock \href {http://dx.doi.org/10.1021/acs.langmuir.0c03162}
  {\path{doi:10.1021/acs.langmuir.0c03162}}.

\bibitem{WolframResearch2012}
{Wolfram Research Inc.}, {Mathematica} (2019).

\bibitem{dosSantos2010}
A.~P. dos Santos, Y.~Levin, {Surface Tensions and Surface Potentials of Acid
  Solutions.}, J. Chem. Phys. 133~(15) (2010) 154107.
\newblock \href {http://dx.doi.org/10.1063/1.3505314}
  {\path{doi:10.1063/1.3505314}}.

\bibitem{vacha2008}
R.~V{\'{a}}cha, D.~Horinek, M.~L. Berkowitz, P.~Jungwirth, {Hydronium and
  Hydroxide at the Interface Between Water and Hydrophobic Media}, Phys. Chem.
  Chem. Phys. 10 (2008) 4675--2980.
\newblock \href {http://dx.doi.org/10.1039/b812223g}
  {\path{doi:10.1039/b812223g}}.

\bibitem{Duignan2015}
T.~T. Duignan, D.~F. Parsons, B.~W. Ninham,
  \href{http://dx.doi.org/10.1016/j.cplett.2015.06.002}{{Hydronium and
  hydroxide at the air-water interface with a continuum solvent model}}, Chem.
  Phys. Lett. 635 (2015) 1--12.
\newblock \href {http://dx.doi.org/10.1016/j.cplett.2015.06.002}
  {\path{doi:10.1016/j.cplett.2015.06.002}}.
\newline\urlprefix\url{http://dx.doi.org/10.1016/j.cplett.2015.06.002}

\bibitem{Tse2015}
Y.-l.~S. Tse, C.~Chen, G.~E. Lindberg, R.~Kumar, G.~A. Voth, {Propensity of
  Hydrated Excess Protons and Hydroxide Anions for the Air-Water Interface}, J.
  Am. Chem. Soc. 137 (2015) 12610--12616.
\newblock \href {http://dx.doi.org/10.1021/jacs.5b07232}
  {\path{doi:10.1021/jacs.5b07232}}.

\bibitem{Mamatkulov2017}
S.~Mamatkulov, C.~Allolio, R.~Netz, D.~J. Bonthuis,
  \href{http://doi.wiley.com/10.1002/anie.201707391}{{Orientation induces
  adsorption of the hydrated proton at the air-water interface}}, Angew. Chemie
  56 (2017) 15846--15851.
\newblock \href {http://dx.doi.org/10.1002/anie.201707391}
  {\path{doi:10.1002/anie.201707391}}.
\newline\urlprefix\url{http://doi.wiley.com/10.1002/anie.201707391}

\bibitem{Marcus2010}
Y.~Marcus, \href{http://iupac.org/publications/pac/82/10/1889/}{{Effect of ions
  on the structure of water}}, Pure Appl. Chem. 82~(10) (2010) 1889.
\newblock \href {http://dx.doi.org/10.1351/PAC-CON-09-07-02}
  {\path{doi:10.1351/PAC-CON-09-07-02}}.
\newline\urlprefix\url{http://iupac.org/publications/pac/82/10/1889/}

\bibitem{Yaminsky1995}
H.~K. Christenson, V.~V. Yaminsky, {Solute Effects on Bubble Coalescence}, J.
  Phys. Chem. 99~(12) (1995) 10420.
\newblock \href {http://dx.doi.org/10.1021/j100025a052}
  {\path{doi:10.1021/j100025a052}}.

\bibitem{Ruckenstein1974}
E.~Ruckenstein, R.~K. Jain, {Spontaneous rupture of thin liquid films}, J.
  Chem. Soc. Faraday Trans. 2 Mol. Chem. Phys. 70 (1974) 132--147.
\newblock \href {http://dx.doi.org/10.1039/F29747000132}
  {\path{doi:10.1039/F29747000132}}.

\bibitem{Tsekov2010}
R.~Tsekov, D.~S. Ivanova, R.~Slavchov, B.~Radoev, E.~D. Manev, A.~V. Nguyen,
  S.~I. Karakashev, {Streaming potential effect on the drainage of thin liquid
  films stabilized by ionic surfactants}, Langmuir 26~(7) (2010) 4703--4708.
\newblock \href {http://dx.doi.org/10.1021/la903593p}
  {\path{doi:10.1021/la903593p}}.

\bibitem{Karakashev2011}
S.~I. Karakashev, R.~Tsekov, {Electro-marangoni effect in thin liquid films},
  Langmuir 27~(6) (2011) 2265--2270.
\newblock \href {http://dx.doi.org/10.1021/la1044656}
  {\path{doi:10.1021/la1044656}}.

\bibitem{Katsir2014a}
Y.~Katsir, A.~Marmur, {Rate of Bubble Coalescence Following Dynamic Approach:
  Collectivity-Induced Specificity of Ionic Effect}, Langmuir 30 (2014)
  13823--13830.

\bibitem{Mundy2009}
C.~J. Mundy, I.~F.~W. Kuo, M.~E. Tuckerman, H.~S. Lee, D.~J. Tobias,
  \href{http://dx.doi.org/10.1016/j.cplett.2009.09.003}{{Hydroxide Anion at the
  Air-Water Interface}}, Chem. Phys. Lett. 481~(1-3) (2009) 2--8.
\newblock \href {http://dx.doi.org/10.1016/j.cplett.2009.09.003}
  {\path{doi:10.1016/j.cplett.2009.09.003}}.
\newline\urlprefix\url{http://dx.doi.org/10.1016/j.cplett.2009.09.003}

\bibitem{Baer2014a}
M.~D. Baer, I.-F.~W. Kuo, D.~J. Tobias, C.~J. Mundy,
  \href{http://www.ncbi.nlm.nih.gov/pubmed/24762096}{{Toward a Unified Picture
  of the Water Self-Ions at the Air-Water Interface: A Density Functional
  Theory Perspective.}}, J. Phys. Chem. B 118~(28) (2014) 8364--8372.
\newblock \href {http://dx.doi.org/10.1021/jp501854h}
  {\path{doi:10.1021/jp501854h}}.
\newline\urlprefix\url{http://www.ncbi.nlm.nih.gov/pubmed/24762096}

\bibitem{Uematsu2018b}
Y.~Uematsu, D.~J. Bonthuis, R.~R. Netz,
  \href{https://linkinghub.elsevier.com/retrieve/pii/S2451910318301868}{{Impurity
  effects at hydrophobic surfaces}}, Curr. Opin. Electrochem. (2018)
  10.1016/j.coelec.2018.09.003\href
  {http://dx.doi.org/10.1016/j.coelec.2018.09.003}
  {\path{doi:10.1016/j.coelec.2018.09.003}}.
\newline\urlprefix\url{https://linkinghub.elsevier.com/retrieve/pii/S2451910318301868}

\bibitem{SweetaAkbari2018}
{Sweeta Akbari}, {Abdurahman Hamid Nour}, {Emulsion types, stability mechanisms
  and rheology: A review}, Int. J. Innov. Res. Sci. Stud. 1~(1) (2018) 14--21.

\end{thebibliography}

\end{document}